\documentclass{article}
\usepackage{jheppub} 
\usepackage{graphicx,color}
\usepackage{epstopdf}
\usepackage{amssymb}
\usepackage{amsmath}
\usepackage{subfigure}
\usepackage{epsfig}
\usepackage{float}
\usepackage{bm}
\usepackage[T1]{fontenc} 
\usepackage{array}
\usepackage{hyperref}
\usepackage{here}
\graphicspath{{figures/}}

\def\simge{\mathrel{%
       \rlap{\raise 0.511ex \hbox{$>$}}{\lower 0.511ex \hbox{$\sim$}}}}
\def\simle{\mathrel{
       \rlap{\raise 0.511ex \hbox{$<$}}{\lower 0.511ex \hbox{$\sim$}}}}

\makeatletter

\newcommand*{\doi}[1]{\href{https://doi.org/#1}{doi:\ #1}}
\newcommand{\figcaption}[1]{\def\@captype{figure}\caption{#1}}
\newcommand{\tblcaption}[1]{\def\@captype{table}\caption{#1}}

\newcommand{\Nsite}{N_{\rm \!full}}

\newcommand{\vsub}{V_{\rm \!sub}}

\newcommand{\nape}{n_{\rm \!APE}}

\newcommand{\SU}{$\mathrm{SU}$}

\newcommand{\pde}{\partial}
\renewcommand{\th}{\theta}

\makeatother

\begin{document}
\title{$\th$ dependence of $T_c$ in \SU(2) Yang-Mills theory}
\def\KEK{Theory Center, Institute of Particle and Nuclear Studies, High
Energy Accelerator Research Organization (KEK), Tsukuba 305-0801, Japan}
\def\SOKENDAI{Graduate University for Advanced Studies (SOKENDAI), %
Tsukuba 305-0801, Japan}
\def\Hongo{Department of Physics, University of Tokyo, 
Tokyo 113-0033, Japan}
\def\IPMU{Kavli Institute for the Physics and Mathematics of the
Universe, University of Tokyo, 
Chiba 277-8583, Japan}
\def\TSQS{Trans-Scale Quantum Science Institute, University of
Tokyo, Tokyo 113-0033, Japan}
\def\CDDD{Center for Data-Driven Discovery, Kavli IPMU, 
University of Tokyo, 
Chiba 277-8583, Japan}
\author[a,b]{Norikazu Yamada,}
\emailAdd{norikazu.yamada@kek.jp}
\author[c,d,e,f]{Masahito Yamazaki,}
\emailAdd{masahito.yamazaki@ipmu.jp}
\author[a,b]{Ryuichiro Kitano}
\emailAdd{ryuichiro.kitano@kek.jp}
\affiliation[a]{\KEK}
\affiliation[b]{\SOKENDAI}
\affiliation[c]{\Hongo}
\affiliation[d]{\IPMU}
\affiliation[4]{\TSQS}
\affiliation[f]{\CDDD}

\date{\today}

\abstract{
We present an exploratory study to determine the confinement-deconfinement transition temperature at finite $\theta$, $T_c(\theta)$, for the 4d \SU(2) pure Yang-Mills theory. Lattice numerical simulations are performed on three spatial sizes $N_S=24$, $32$, $48$ with a fixed temporal size $N_T=8$. We introduce a non-zero $\theta$-angle by the sub-volume method to mitigate the sign problem. By taking advantage of the universality in the second order phase transition and the Binder cumulant of the order parameter, the $\theta$-dependence of $T_c$ is determined to be $T_c(\theta)/T_c(0)=1-0.16(2)\,(\theta/\pi)^2-0.03(4)\,(\theta/\pi)^4$ up to $\theta\sim 0.9\,\pi$. The reliability of the extrapolations in the sub-volume method is extensively checked. We also point out that the temperature dependence of the topological susceptibility should exhibit a singularity with the exponent for the specific heat.}

\maketitle

\section{Introduction}

The Yang-Mills (YM) theory is one of the best-studied quantum field theories and 
play central roles in the standard model of particle physics.
The pure YM theory corresponds to the static limit of fermions and provides avenues to explore nonperturbative dynamics.

The goal of this paper is to discuss the $\th$-$T$ phase diagram, in particular $\theta$-angle dependence of the confinement-deconfinement transition temperature $T_c$, for the 4d \SU($2$) pure YM theory.

Our lattice numerical simulations are motivated by some previous studies.
In our previous paper~\cite{Kitano:2021jho} (in collaboration with R.~Matsudo),
we calculated the $\th$ dependence of the vacuum energy density.
Using the sub-volume method developed to overcome the sign problem, the parity symmetry in the vacuum is found to be spontaneously broken at $\th=\pi$~\cite{Kitano:2021jho}.
A crosscheck of this finding is possible by investigating whether $T_c(\pi)>0$ or not~\cite{Gaiotto:2017yup,Kitano:2017jng}, where $T_c(\th)$ is the confinement-deconfinement transition temperature at a finite $\th$.

The case of the \SU($2$) gauge group is of particular interest, since we expect some qualitative differences from \SU($N_c$) theories with $N_c\ge 3$. The $\th$ dependence of the critical temperature has been numerically studied in the vicinity of $\th=0$ for \SU($N_c$) theories with $N_c\ge 3$~\cite{DElia:2012pvq,DElia:2013uaf,Otake:2022bcq,Borsanyi:2022fub,Bonanno:2023hhp}\footnote{For analytical investigations, see, for example, Refs.~\cite{Unsal:2012zj,Poppitz:2012nz,Anber:2013sga}.}, where the system undergoes the first order phase transition.
The outcome is that the dependence can be written as
$T_c(\th)/T_c(0)=1-R\,\th^2+O(\th^4)$ with
$R\sim 0.17/N_c^2$ for ($N_c\ge 3$)~\cite{DElia:2012pvq,DElia:2013uaf,Bonanno:2023hhp}.
Since $R$ increases with $1/N_c^{2}$, $T_c(\th)$ for $N_c=2$ may vanish at $\th\sim O(1)$ and
a gapless theory may emerge at $\th=\pi$,
as for the 2d CP$^{1}$ model~\cite{Haldane:1983ru,Haldane:1982rj,Affleck:1987ch,Shankar:1989ee}
which shares many features with the 4d \SU(2) pure YM theory.\footnote{See \cite{Yamazaki:2017ulc,Yamazaki:2017dra} for more precise correspondence between 4d \SU(2) pure YM theory and a version of the 2d CP$^{1}$ model.}
One needs to be very careful, however, when extrapolating the $N_c\ge 3$
results to $N_c=2$.
For example, the critical point for the \SU(2) YM theory 
is of second order, while those for \SU($N_c\ge 3$) theories
are of first-order. In the analysis of \SU($N_c\ge 3$) theories
$R$ is computed as the ratio of the jump of the topological susceptibility at $T_c(0)$ to the latent heat via
the Clausius-Clapeyron relation, which is available only for the first-order phase transitions, and there is no such jump for the 
second-order transition for the \SU(2) YM theory.
This means that a separate analysis is needed for the \SU(2) theory.

The rest of the paper is organized as follows.
After simulation parameters and the definition of topological charge on the lattice are summarized in sec.~\ref{sec:setup}, we outline the method in sec.~\ref{sec:method}.
The re-weighting method is used to introduce the $\th$ term. In order to mitigate the sign problem, we apply the sub-volume method in which the re-weighting is applied only to a sub-volume of the whole lattice at the intermediate step. The sub-volume results are then extrapolated to the full-volume.
We use this method to calculate the Binder cumulant of the order parameter at finite $\th$ to identify the critical point.
In sec.~\ref{sec:test-of-critical-behaviors}, we verify the universality around the critical point by demonstrating that the critical behaviors in the 4d \SU(2) gauge theory are well described by the critical exponents in the 3d Ising model.
The universality between these two theories is extensively used in this work.
The numerical results are presented in sec.~\ref{sec:results} including $T_c(\th)/T_c(0)$ as a function of $\th$.
An alternative analysis is given to estimate the systematic uncertainty originating from the extrapolation.
In the sub-volume method, it is crucial that the extrapolation to the full-volume is under good control.
In sec.~\ref{sec:check}, to confirm the reliability of the extrapolation, we see whether the results for the Binder cumulant at different values of $\th$ and the lattice coupling approximately form a single curve as they should in the second order phase transition.
In sec.~\ref{sec:discussion}, based on analytic considerations it is pointed out that the $\th$ dependence is determined by the ratio of the critical behaviors of thermodynamic quantities, and the topological susceptibility $\chi_{t}$ has to have a singular $T$ dependence with the same exponent as does the specific heat.
Finally, concluding remarks are presented in sec~\ref{sec:summary}.

\section{Lattice set-up}
\label{sec:setup}

\subsection{Parameters}\label{subsec:parameters}

The \SU(2) lattice gauge action at $\theta=0$ is given by
\begin{align}
 S_g = 6\,\beta_g\,\Nsite\,\left\{c_0(1-W_P)+2\,c_1(1-W_R)\right\}
 \ ,
 \label{eq:action}
\end{align}
where $\beta_g=4/g^2$ is the inverse lattice gauge coupling.
The $1\times 1$ plaquette and the $1\times 2$ rectangle are constructed
from \SU(2) link variables, and $W_P$ and $W_R$ are those averaged over
four-dimensional lattice sites ($\Nsite=N_S^3\times N_T$) and all
possible directions, respectively.
The coefficients $c_0$ and $c_1$ satisfying $c_0=1-8c_1$ are the improvement
coefficients, and we take the tree-level Symanzik improved action,
$c_1=-1/12$~\cite{Weisz:1982zw}.
The calculations are carried out on three spatial lattice sizes ($N_S=$24, 32, 48) with the temporal
size fixed to $N_T=8$.
The critical value of $\beta_g$ for $N_T=8$ at $\theta=0$, corresponding to the critical temperature $T_c(0)$, is known to be $\beta_g^{\rm crit}=1.920(5)$~\cite{Cella:1993ic,Giudice:2017dor}.

The gauge ensembles used in this work were generated by the Hybrid Monte Carlo method with periodic boundary conditions
in all directions and stored in every ten trajectories after thermalization in \cite{Yamada:2024vsk}.
The statistics of some ensembles have been increased since \cite{Yamada:2024vsk}.
Statistical errors are estimated by the single-elimination jack-knife method with the bin size of 1,000 configurations.
We summarize in Tab.~\ref{tab:parameters} simulation parameters including $\beta_g$, the corresponding $T/T_c(0)$, the lattice size, and the statistics.
\begin{table}[tbh]
 \begin{center}
 \begin{tabular}{|r|rrrrrrr|}
 \hline
  $\beta$ & 1.87 & 1.88 & 1.89 & 1.90 & 1.91 & 1.92 & 1.93\\
  $T/T_c(0)$ & 0.85 & 0.88 & 0.91 & 0.94 & 0.97 & 1.00 & 1.03\\
 \hline
  $\!\!N_S$=24~&120&110& 90& 80& 63& 50& 20\\
            32 &20 & 40&170& 60& 60& 57& 50\\
            48 &10 &   & 20& 30& 60& 38& 35\\
  \hline
 \end{tabular}
 \caption{
  The simulation parameters and statistics.
  The numbers of configurations are shown in units of 1,000.
  The temporal size $N_T$ is fixed to 8.
  The normalized temperature $T/T_c(0)$ is obtained by fitting the data presented in~\cite{Giudice:2017dor}
  (see Fig.~\ref{fig:beta-vs-tovtc}).
  }
 \label{tab:parameters}
 \end{center}
\end{table}

\begin{figure}[th]
  \begin{center}
  \begin{tabular}{c}
  \includegraphics[width=0.7 \textwidth]{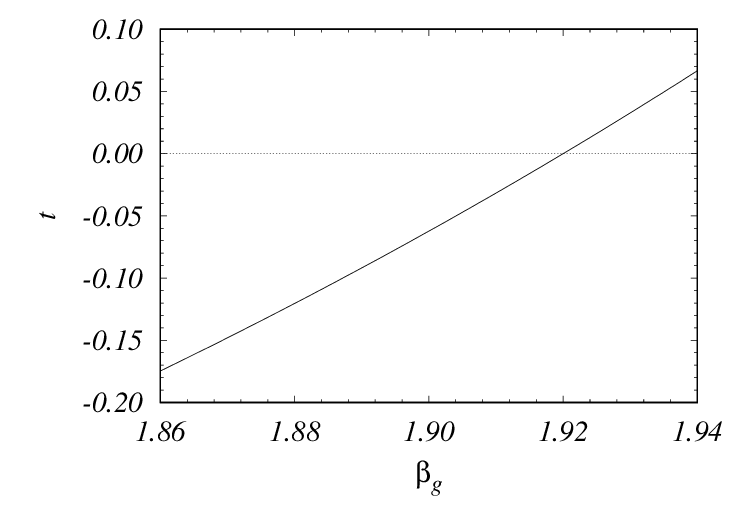}\\[-3ex]
  \end{tabular}
  \end{center}
 \caption{The correspondence between $\beta_g$ and the reduced temperature $t$ at $\th=0$.
 The uncertainty of the fitted curve is small within the range shown here and neglected in the following
 analysis as it does not affect the conclusion.
 }
\label{fig:beta-vs-tovtc}
\end{figure}
Below we will often quote the reduced temperature $t:=T/T_c-1$.
The correspondence between $t$ and $\beta_g$ at $\th=0$ is shown in Fig.~\ref{fig:beta-vs-tovtc}.

\subsection{Topological charge density and smearing}
\label{subsec:topological-charge-density}

In this work, non-zero values of $\th$ are introduced through the re-weighting
on a sub-volume, which requires the local values of the topological charge $q(x)$.
The conceptually cleanest way to measure $q(x)$ would be to use the
eigenvectors of the overlap Dirac operator~\cite{Hasenfratz:1998ri}.
However, this method is not realistic for high-statistics and large-volume lattice calculations because the computational cost is too large.
Instead, we use the bosonic definition, where the topological term
$G(x)\tilde{G}(x)$ is directly constructed from the \SU(2) link variables.
We adopt the five-loop improved topological charge operator introduced in~\cite{deForcrand:1997esx}.

The topological charge distribution thus obtained is contaminated by lattice artifacts, which are eliminated by smoothing the configurations.
The smoothing procedure, however, may simultaneously deform physically legitimate topological excitations.
To restore original topological information from deformed one, we follow the study in~\cite{Kitano:2020mfk} and extrapolate the data with different smearing steps to zero smearing.
Among the several mutually consistent methods often used in the
literature~\cite{Bonati:2014tqa,Alexandrou:2015yba,Alexandrou:2017hqw},
we apply the APE smearing~\cite{Albanese:1987ds} to produce smeared configurations, on which the local topological charge $q(x,\nape)$ is calculated.
This procedure and the parameters for the APE smearing are precisely the same
as those in~\cite{Kitano:2020mfk,Kitano:2021jho,Yamada:2024vsk}.

A local topological charge is calculated at every five smearing steps,
{\it i.e.}\ $n_{\rm APE}=$ 0, 5, 10, $\cdots, 50$ and is uniformly shifted as $q(x,\nape)\to q(x,\nape)+\epsilon$
such that the global topological charge $Q:=\sum_{x\in \Nsite}q(x,\nape)$ takes the integer value closest to the original one.
Topology fluctuates appropriately in all ensembles studied here~\cite{Yamada:2024vsk}.

\section{Method}
\label{sec:method}
\subsection{Binder cumulant}
\label{subsec:binder}

In order to determine the curve $T_c(\th)$ on the $\th$-$T$ plane, the confinement-deconfinement
transition point needs to be identified at a finite $\th$.
For this purpose, we use the fourth-order Binder cumulant~\cite{Binder:1981sa} $B_4$, which is constructed
from the cumulants of the order parameter, {\it i.e.} the Polyakov loop $\omega$, as
\begin{align}
 B_4(\beta_g,\th)
&= 1-\frac{\langle \omega^4 \rangle_{\beta_g,\th} }{3\,
      \langle \omega^2 \rangle_{\beta_g,\th}^2}
 \label{eq:binder}
 \,,\\
   \omega
&= \frac{1}{N_S^3}
   \sum_{x\in N_S^3}\frac{1}{N_c}{\rm Tr}
   \left[\Pi_{n_t=1}^{N_T}U_{4}(x,n_t)\right]
\label{eq:poly}
\,,
\end{align}
where $U_4$ denotes a link variable in the temporal direction and
the subscripts $\beta_g$ and $\th$ the lattice parameters at which
the expectation value is estimated.
The Binder cumulant has been widely used for the identification of critical points in various systems.
Importantly this quantity is known to be less affected by finite volume corrections, and
hence is suitable for the calculations relying on relatively small lattices.
Below the critical $\beta_g$ ($\beta_g^{\rm crit}$), the distribution of $\omega$
approaches the Gaussian toward the large volume limit and $B_4$ approaches
zero, while above $\beta_g^{\rm crit}$, $B_4$ approaches $2/3$ in the same limit.

The second-order phase transitions in 4d \SU(2) pure YM theory and the 3d Ising model share the critical value of $B_4$ ($B_4^{\rm crit}$)
as we will explicitly verify later.
In order to identify the critical point, we calculate $B_4$ at various
$\beta_g$ and $\th$ and look for the point where $B_4$ is equal to $B_4^{\rm crit}$.
Of many works having determined $B_4^{\rm crit}$ in 3d Ising model, we refer $B_4^{\rm crit}=0.466$~\cite{Hasenbusch:1998gh}
whose error is smaller than the last digit.

\subsection{Reweighting on sub-volume}
\label{subsec:reweighting}

Non-zero $\th$ is introduced through the re-weighting method.
Inspired by~\cite{Kitano:2021jho,Yamada:2024vsk}, in order to mitigate the sign problem, we propose to apply the re-weighting only to a sub-volume at the intermediate step as follows.
Using the local topological charge $q(x,\nape)$ described in sec.~\ref{subsec:topological-charge-density}, we define
the sub-volume topological charge by
\begin{align}
& Q_l(\nape) = \displaystyle\sum_{x\in \vsub} q(x,n_{\rm APE})
\,,
 \label{eq:qsub}
\end{align}
where $\vsub=l^3\times N_T$ and $N_T\le l\le N_S$.
Using $Q_l$, we calculate two quantities,
\begin{align}
   \langle {\omega_s}^n \rangle_{\beta_g,\th,\nape,l}\
&= {\langle {\omega_l}^n \cos(\th Q_l(\nape))\rangle_{\beta_g}
    \over
    \langle \cos(\th Q_l(\nape))\rangle_{\beta_g}}\,,
\\
   \langle {\omega_f}^n \rangle_{\beta_g,\th,\nape,l}\
&= {\langle {\omega_{\rm full}}^n \cos(\th Q_l(\nape))\rangle_{\beta_g}
    \over
    \langle \cos(\th Q_l(\nape))\rangle_{\beta_g}}
\,,
\end{align}
where $\omega_{\rm full}$ is $\omega$ itself and $\omega_l$ is the one replacing $N_S^3$
by $l^3$ in~\eqref{eq:poly} which covers the same sub-volume as does $Q_l$.
The expectation values on the right-hand side of the above equations are evaluated
on gauge ensembles generated with $\th=0$.
$B_{4,i}$ ($i=s$ or $f$) is then obtained by
\begin{align}
 B_{4,i}(\beta_g,\th,\nape,l)
&= 1-{\langle {\omega_i}^4 \rangle_{\beta_g,\th,\nape,l} \over3\,
      \langle {\omega_i}^2 \rangle_{\beta_g,\th,\nape,l}^2}\,.
\end{align}
Notice that the only difference between $B_{4,s}$ and $B_{4,f}$ is the volume
over which the Polyakov loop is averaged; they thus coincide in the limit of $l\to N_S$.

We obtain
$B_4(\beta_g,\th)=B_{4,i}(\beta_g,\th,0,N_S)$ by taking
the following limits,
\begin{align}
   B_4(\beta_g,\th)
=& \lim_{\nape\to 0}\lim_{l\to N_S}
   B_{4,i}(\beta_g,\th,\nape,l)
   \label{eq:extrapolation-B4}
\,.
\end{align}
Note that the sub-volume data are extrapolated to the full-volume but
not infinitely large volume because the analysis of the Binder cumulant requires those at finite full-volume.
Therefore, the extrapolation is more tractable than those in~\cite{Kitano:2021jho,Yamada:2024vsk}.
Furthermore, the simultaneous fit using two quantities $B_{4,s}$ and
$B_{4,f}$ sharing the same limit is expected to make the fit more stable than fitting only either one of them.
With the full use of translational invariance, the results on smaller sub-volumes are statistically more accurate.
While we can exchange the order of the two limits, we have chosen the ordering as in the equation above because it gives slightly better accuracy.

\section{Test of universality}
\label{sec:test-of-critical-behaviors}

\begin{figure}[ht]
  \begin{center}
  \begin{tabular}{cc}
  \includegraphics[width=0.5 \textwidth]{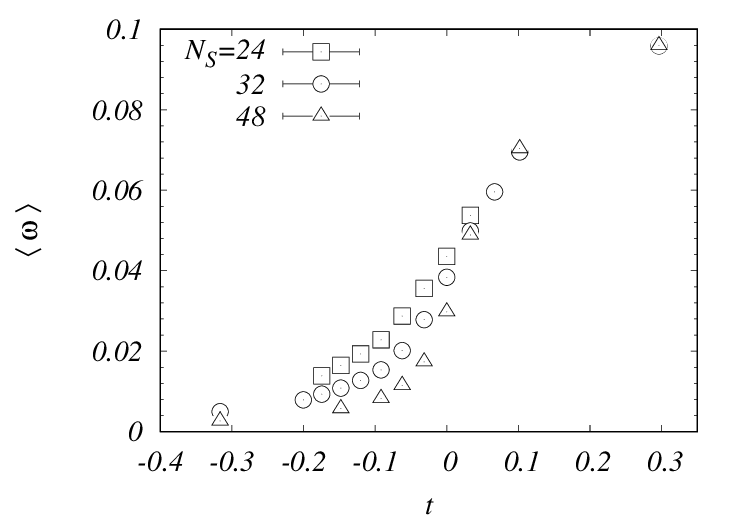} &
  \includegraphics[width=0.5 \textwidth]{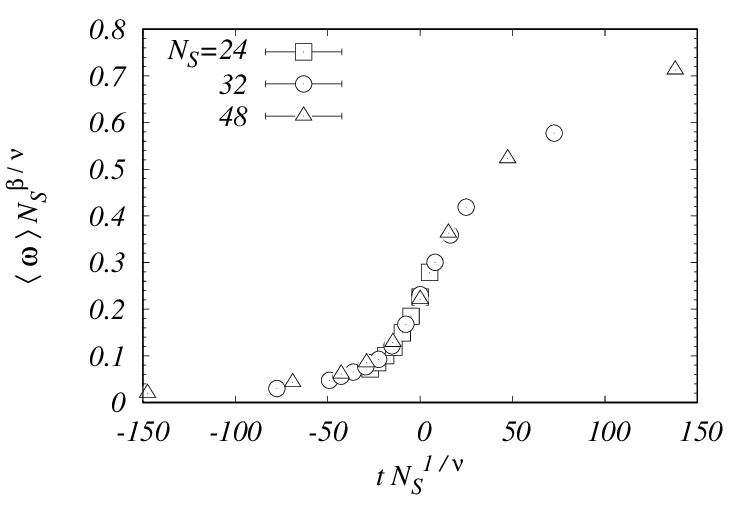}\\
  \includegraphics[width=0.5 \textwidth]{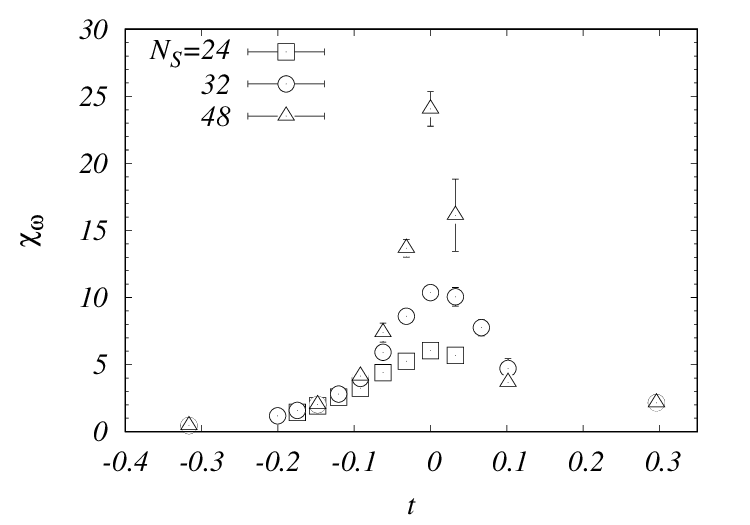} &
  \includegraphics[width=0.5 \textwidth]{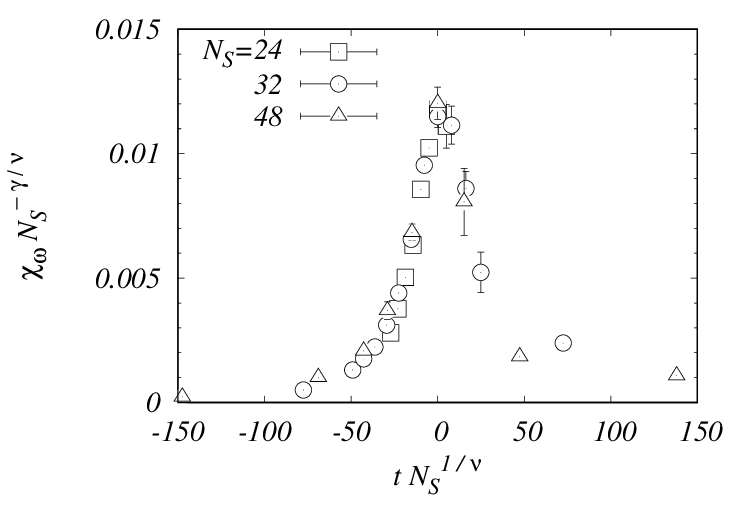}\\
  \includegraphics[width=0.5 \textwidth]{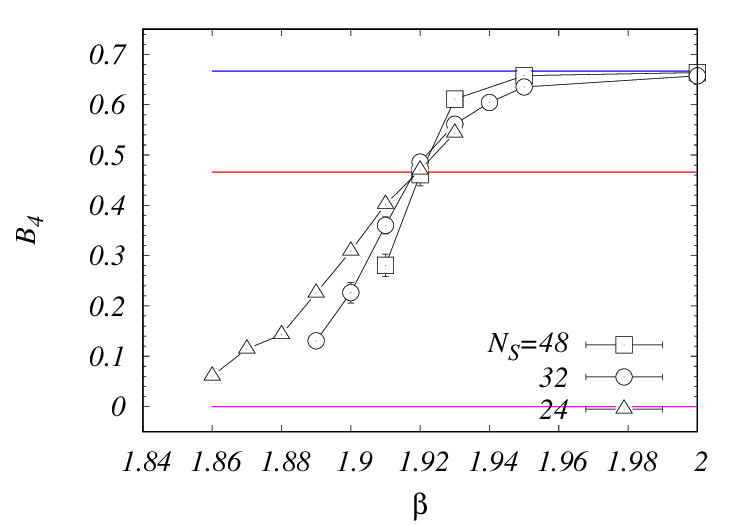}&
  \includegraphics[width=0.5 \textwidth]{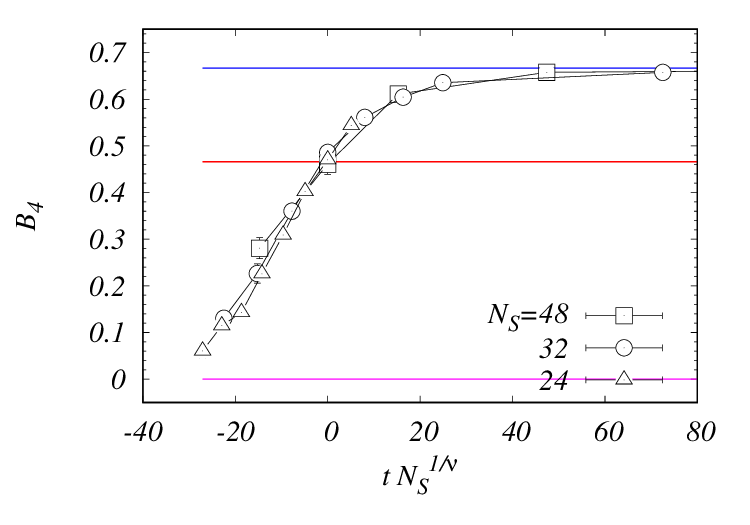}
  \end{tabular}
  \end{center}
 \caption{
 The raw data (left) and the rescaled data (right) of the Polyakov loop (top),
 its susceptibility (middle) and $B_4$ (bottom) at $\th=0$, where $\beta_g^{\rm crit}=1.92$
 and the critical exponents in the 3d Ising model are used.
 The data with $\beta_g<1.87$ and $\beta_g>1.93$ are from pilot calculations at $\th=0$.
 }
\label{fig:scaling-plot}
\end{figure}

Before showing the main results, we verify that our ensembles indeed yield critical behaviors
expected from the universality of second-order phase transition~\cite{Svetitsky:1982gs}.
To be more precise, we compare the numerical results for the expectation
value of the Polyakov loop $\langle\omega\rangle$ and its
susceptibility ($\chi_\omega= N_S^3\left(   \langle \omega^2\rangle - \langle |\omega|\rangle^2\right)$)
with their finite size scaling relations~\cite{Amit:1984ms},
\begin{align}
\langle\omega\rangle
&\approx N_S^{-\beta/\nu}g_1(N_S^{1/\nu}t)
 \label{eq:d1fdh1}
\,,\\
\chi_{\omega}
&\approx N_S^{\gamma/\nu}g_2(N_S^{1/\nu}t)
 \label{eq:d2fdh2}
\,,
\end{align}
where $t$ is the reduced temperature defined previously and $g_1$ and $g_2$ are unknown regular functions.
With the critical exponents $\beta=0.3265(3)$, $\gamma=1.2372(5)$ and $\nu=0.6301(4)$
for the 3d Ising model~\cite{Pelissetto:2000ek},
the temperature dependence of $\langle\omega\rangle$ (top) and $\chi_{\omega}$ (middle)
obtained at $\th=0$ are rescaled as shown in Fig.~\ref{fig:scaling-plot}, where the raw  (left) and rescaled (right) data are shown for comparison neglecting the uncertainties in the exponents.
The rescaled data lie on a single curve in either plot.
For further consistency checks on the critical behaviors and the
determination of the critical temperature, see, for example, 
\cite{Engels:1989fz,Cella:1993ic,Engels:1995em,Engels:1998nv,Giudice:2017dor}.

The fourth-order Binder cumulant provides another consistency check of the universality.
According to the finite size scaling analysis~\cite{Binder:1981sa}, $B_4$ can be expressed at the leading order as
\begin{align}
 B_4(t,1/N_S)
=& {3\,\langle \omega^2 \rangle^2
    -  \langle \omega^4 \rangle
   \over
   3\,\langle \omega^2 \rangle^2}
\approx - {1\over 3}{g_4(N_S^{1/\nu}t)
                \over
                \left(g_2(N_S^{1/\nu}t)\right)^2}
\label{eq:binder-0}
\,,
\end{align}
where $g_4$ is an unknown regular function.
Note that at $t=0$, $B_4$ is independent of $N_S$ and estimated to be $B_4^{\rm crit}=0.466$
in the 3d Ising model~\cite{Hasenbusch:1998gh}.
The raw and rescaled data of $B_4$ calculated at $\th=0$ on three
different volumes are shown at the bottom of Fig.~\ref{fig:scaling-plot},
where three horizontal lines denote the asymptotic values for the low-$T$ (0) and
high-$T$ (2/3) regions in the large volume limit and the critical value (0.466).
In the plot of raw data (left), the three curves intersect at a single point
$(\beta_g,\,B_4)\approx (1.92, 0.466)$, while in the rescaled one (right) they approximately collapse
to a single curve.
In the following analysis, we take the full advantage of the universality between the 4d \SU(2) YM theory and the 3d Ising model.

\section{Numerical results}\label{sec:results}

Let us now present our numerical results.

\begin{figure}[t]
  \begin{center}
  \begin{tabular}{cc}
  \includegraphics[width=0.5 \textwidth]{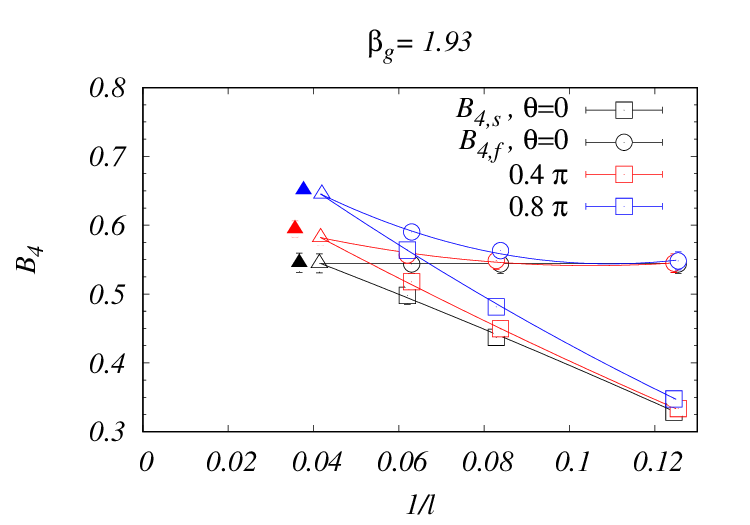}&
  \includegraphics[width=0.5 \textwidth]{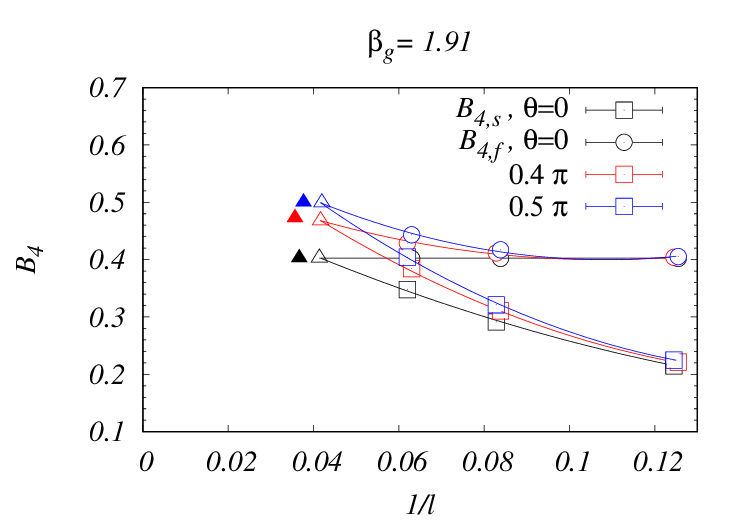}\\
  \includegraphics[width=0.5 \textwidth]{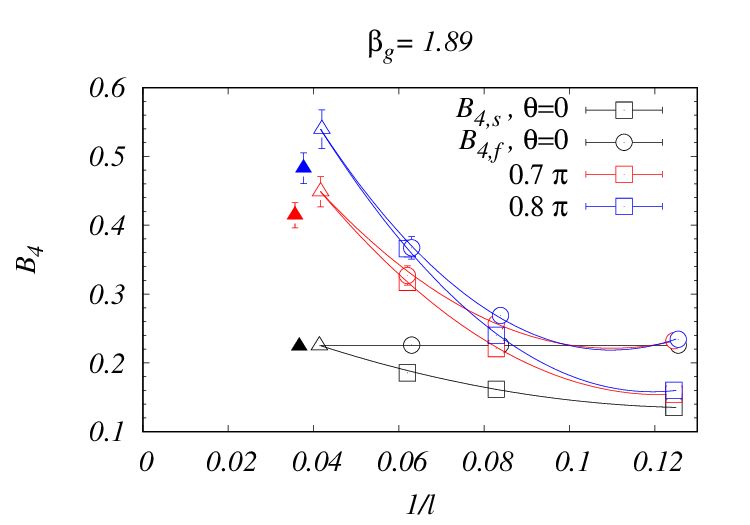}&
  \includegraphics[width=0.5 \textwidth]{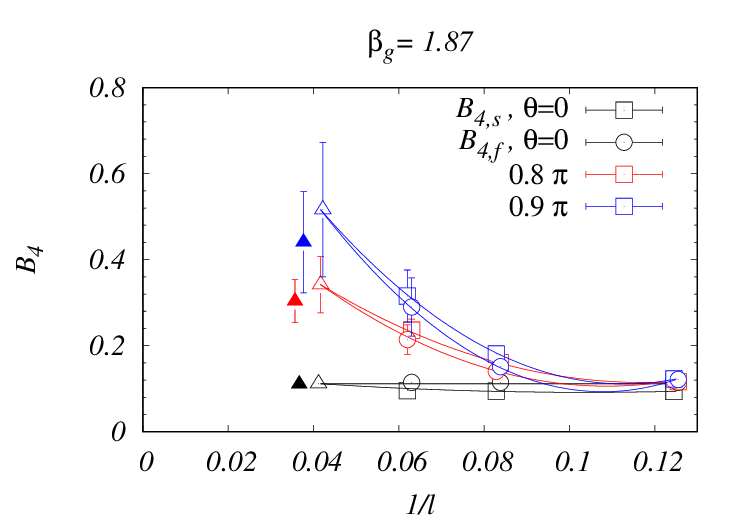}
  \end{tabular}
  \end{center}
 \caption{$l$ dependence of $B_4$ at $n_{\rm APE}=30$ for $N_S=24$
 lattices.
 The data points are slightly shifted in the horizontal direction for clarity.
 The same shift is applied in all the following figures.}
 \label{fig:l-bdc-24}
\end{figure}

\begin{figure}[t]
  \begin{center}
  \begin{tabular}{cc}
  \includegraphics[width=0.5 \textwidth]{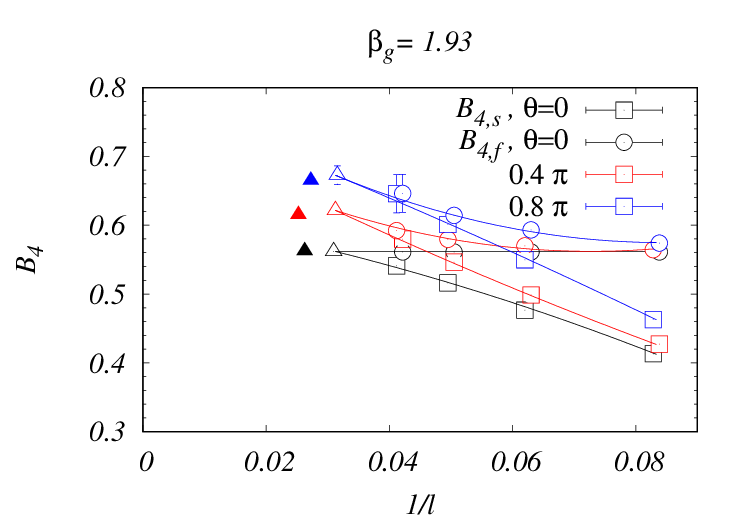}&
  \includegraphics[width=0.5 \textwidth]{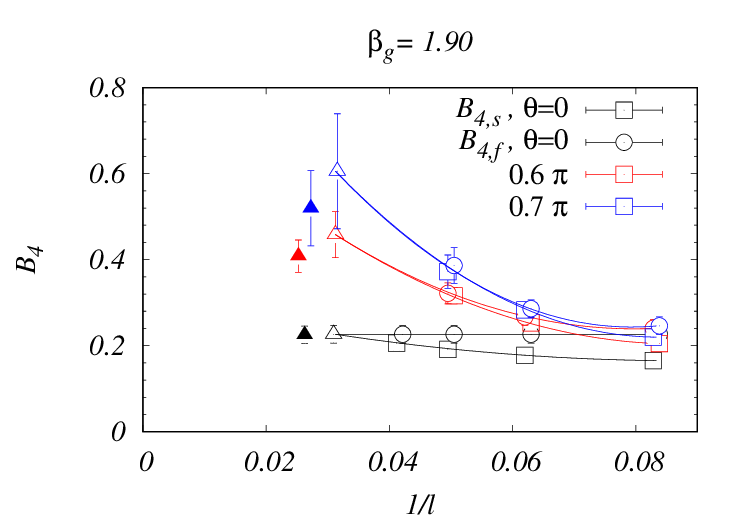}
  \end{tabular}
  \end{center}
 \caption{$l$ dependence of $B_4$ at $n_{\rm APE}=30$ for $N_S=32$
 lattices. }
 \label{fig:l-bdc-32}
\end{figure}

\begin{figure}[t]
  \begin{center}
  \begin{tabular}{cc}
  \includegraphics[width=0.5 \textwidth]{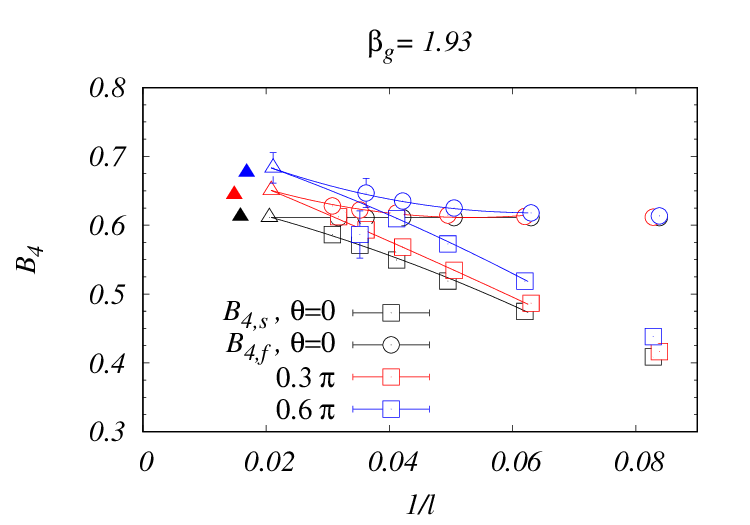}&
  \includegraphics[width=0.5 \textwidth]{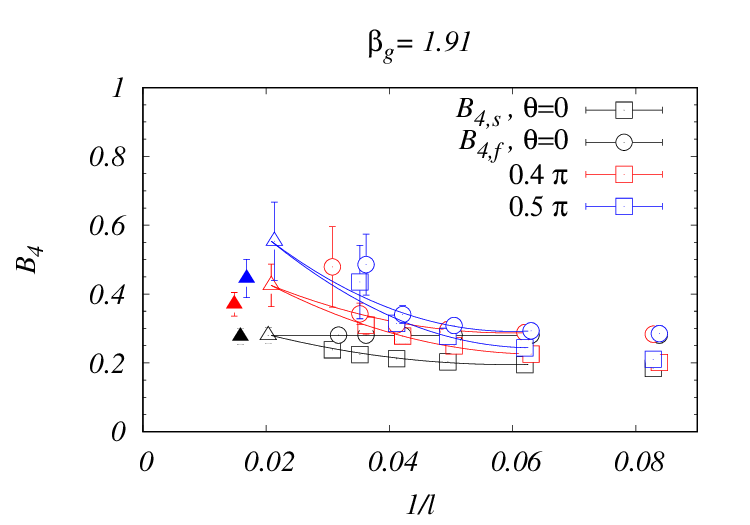}
  \end{tabular}
  \end{center}
 \caption{$l$ dependence of $B_4$ at $n_{\rm APE}=30$ for $N_S=48$
 lattices. }
 \label{fig:l-bdc-48}
\end{figure}

Following \eqref{eq:extrapolation-B4}, the sub-volume results are first
extrapolated to the full-volume.
Examples at $\nape =30$ on $N_S=24$, 32, 48 lattices are shown in
Figs.~\ref{fig:l-bdc-24},~\ref{fig:l-bdc-32} and~\ref{fig:l-bdc-48}, respectively.
On each plot we explicitly show the extrapolations at $\th=0$ and two other values of $\th$.

The value of $B_4$ at $l=N_S$ is finite and hence a simple polynomial fit is expected to work.
We simultaneously fit $B_{4,s}$ and $B_{4,f}$ to a quadratic function of $1/l$ except for the data of $B_{4,f}$ at $\th=0$, which do not depend on $l$ by definition and are fitted to a constant.
In the extrapolation it is preferable to use the data on the sub-volume close to the full-volume, {\em i.e.} $l\sim N_S$.
However, as $\th$ or $l$ increases, the data tends to be noisy or even unavailable due to the sign problem.
By studying the general tendency in the $l$ dependence, we have chosen the fit range to be $l \ge N_S/3$ to keep the number of data points.
The validity of the functional form and the fit range chosen here is not guaranteed a priori.
Thus, it will be checked empirically in the later parts of this paper.

Re-weighting procedure sometimes ends up with unreasonably large statistical uncertainty, and including such data in the analysis may bring in instability.
To avoid such complications and make the fits stable, we only use the data for $B_{4,i}(\beta_g,\th,\nape,l)$ with 30\% or better accuracy.

The data with small $l$ depends weakly on $\th$ since $Q_l$ is small and
the re-weighting factor $\cos(\th Q_l)$ is approximately one.
It is seen that $B_{4,i}(\beta_g,\th,\nape,l)$ increases with $\th$, which
immediately indicates that $\beta_g^{\rm crit}$ decreases with $\th$.
At the same time, as $\th$ increases, the $l$ dependence tends to be large and so does the resulting uncertainty.
In general, the statistical error increases with $\th$ more rapidly at smaller $\beta_g$.

In the sub-volume method, it is not easy to estimate the systematic uncertainty associated with the extrapolation because there are only a few degrees of freedom in changing the fit range.
In this work, to get a rough estimate for the size of the systematic uncertainty, we also performed the simultaneous {\em linear} fit omitting the data with the smallest $l$, whose result is plotted to the left of the main result in each case (filled symbols).

\begin{figure}[t]
  \begin{center}
  \begin{tabular}{cc}
  \includegraphics[width=0.5 \textwidth]{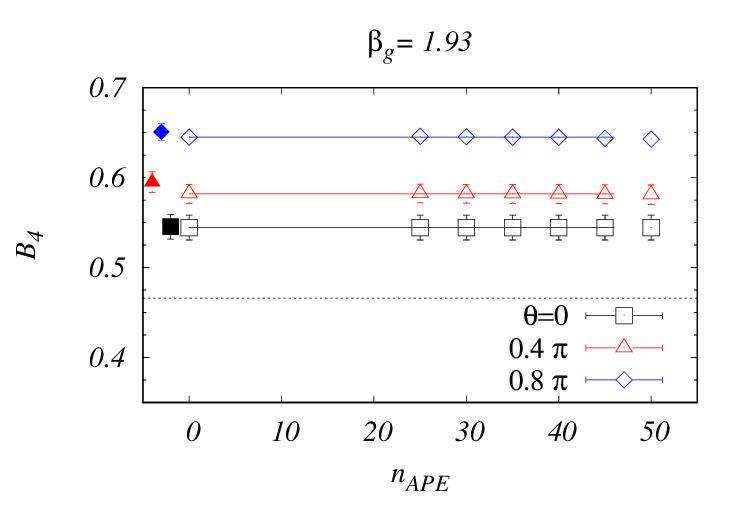}&
  \includegraphics[width=0.5 \textwidth]{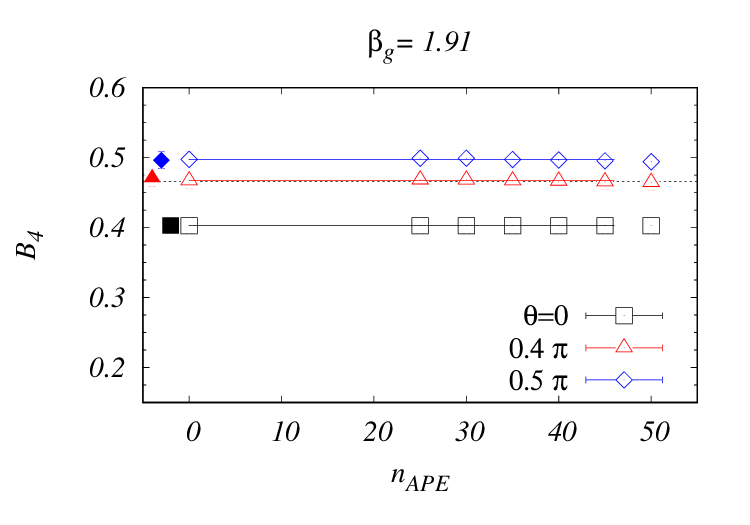}\\
  \includegraphics[width=0.5 \textwidth]{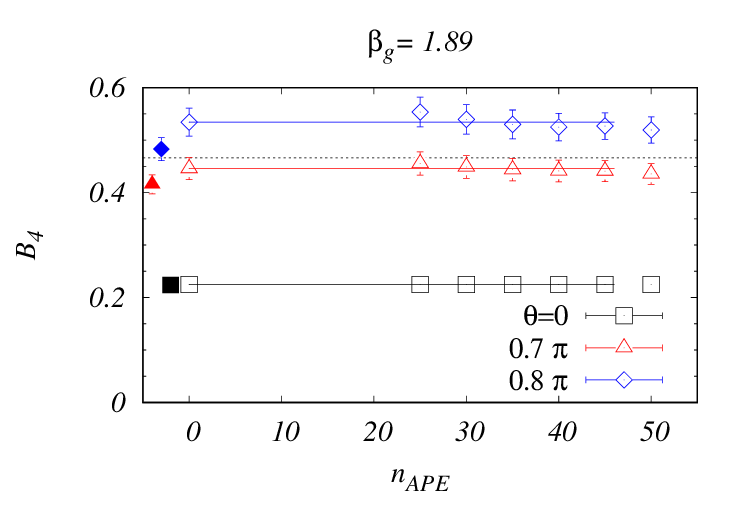}&
  \includegraphics[width=0.5 \textwidth]{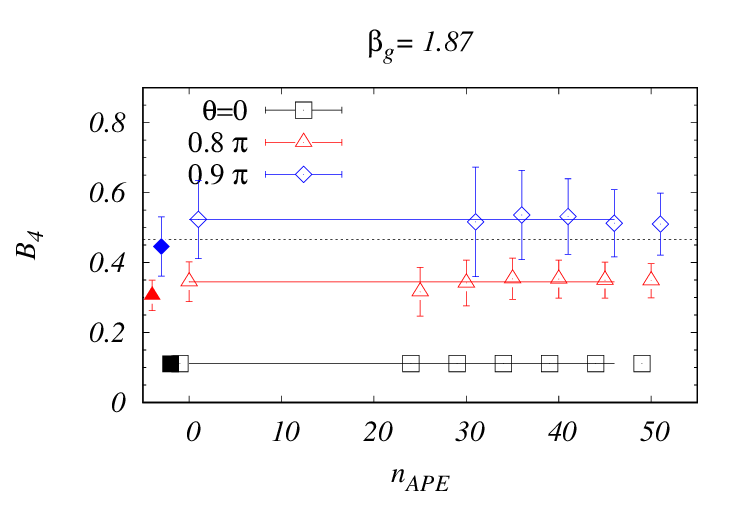}
  \end{tabular}
  \end{center}
 \caption{$n_{\rm APE}$ dependence of $B_4$ for $N_S=24$ lattices. }
 \label{fig:nape-bdc-24}
\end{figure}

\begin{figure}[ht]
  \begin{center}
  \begin{tabular}{cc}
  \includegraphics[width=0.5 \textwidth]{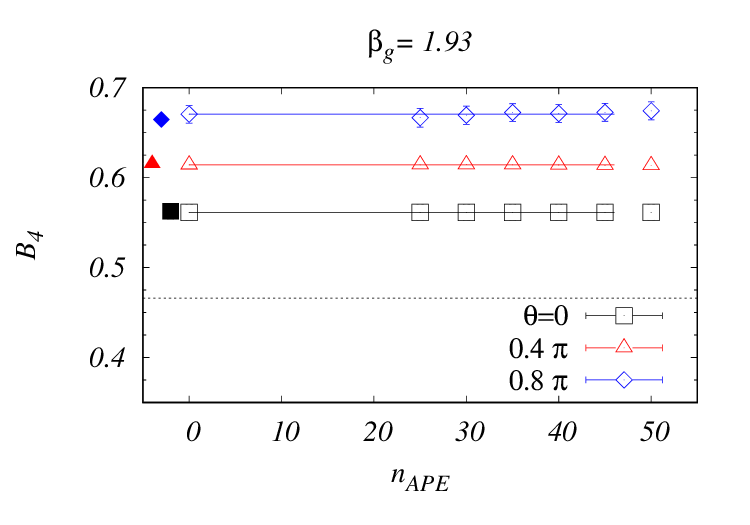}&
  \includegraphics[width=0.5 \textwidth]{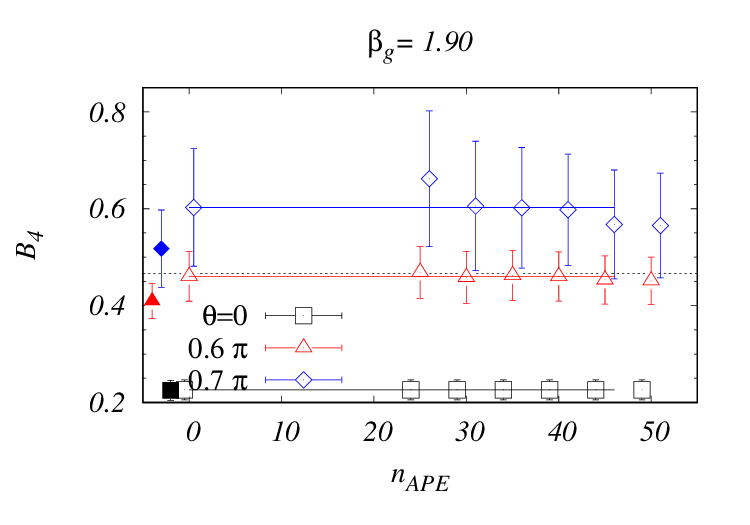}
  \end{tabular}
  \end{center}
 \caption{$n_{\rm APE}$ dependence of $B_4$ for $N_S=32$ lattices. }
 \label{fig:nape-bdc-32}
\end{figure}

\begin{figure}[ht]
  \begin{center}
  \begin{tabular}{cc}
  \includegraphics[width=0.5 \textwidth]{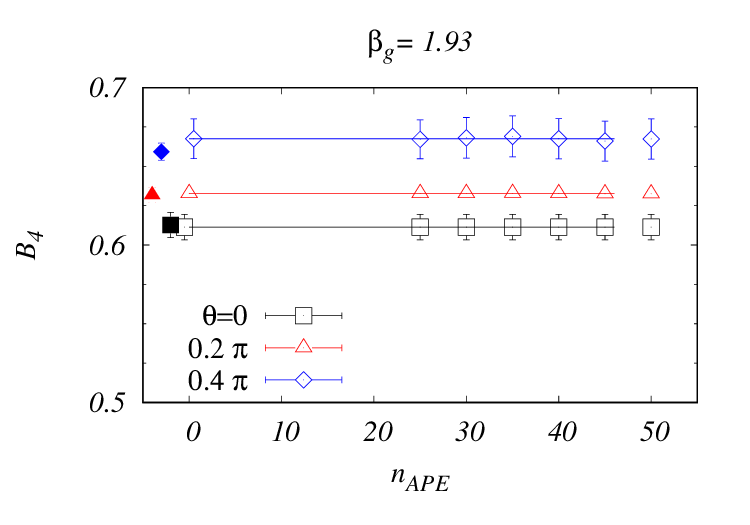}&
  \includegraphics[width=0.5 \textwidth]{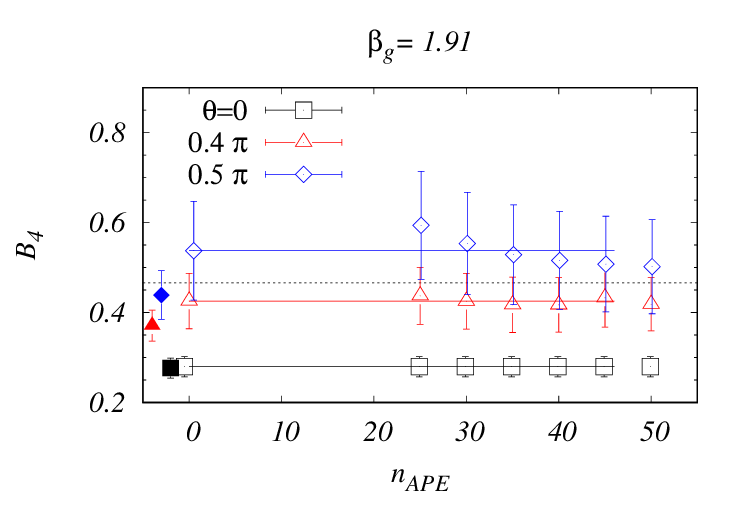}
  \end{tabular}
  \end{center}
 \caption{$n_{\rm APE}$ dependence of $B_4$ for $N_S=48$ lattices.}
 \label{fig:nape-bdc-48}
\end{figure}

The full volume results $B_4(\beta_g,\th,\nape,N_S)$ thus obtained are next extrapolated to $\nape=0$ at each $\th$.
As shown in Figs.~\ref{fig:nape-bdc-24},~\ref{fig:nape-bdc-32} and~\ref{fig:nape-bdc-48}, no $\nape$-dependency is observed when the data are relatively accurate, and importantly we observe some hint of possible $\nape$-dependency only when the data are less accurate.
This observation suggests that the true value of $B_4$ is independent of $\nape$, and any such dependency is due to incomplete extrapolations.
Thus, we fit the data to a constant, taking into account the correlation among the data.
The fit range is chosen to be $25\le\nape\le 45$ in all fits, and the fit range dependence turns out to be smaller than the resulting uncertainty at $\nape=0$.

The critical value for $B_4$ ($B_4^{\rm crit}=0.466$) is indicated in the figures as the horizontal dotted line.
In the plots, in addition to the data with $\th=0$, we show those close to $B_4^{\rm crit}$ if available.
We apply the same analysis to the results from the linear sub-volume extrapolation, and the results are plotted to the left of the main result (filled symbols).
In general, the two results turn out to be consistent with each other within the statistical uncertainty.

\begin{figure}[htb]
  \begin{center}
  \begin{tabular}{cc}
  \includegraphics[width=0.5 \textwidth]{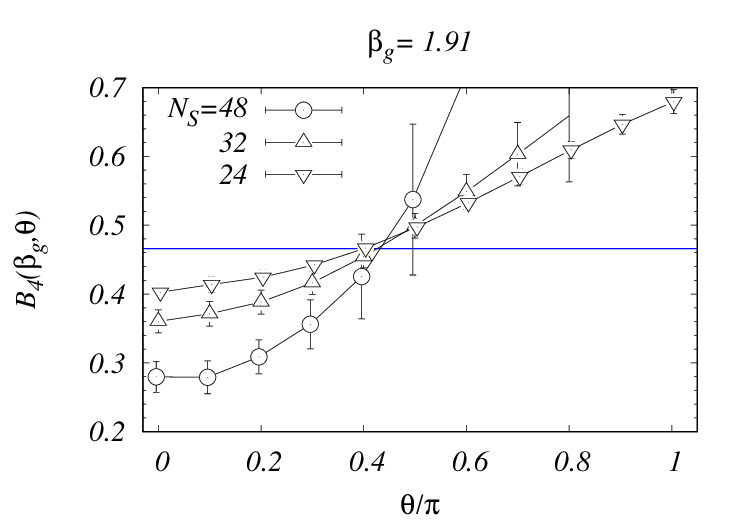}&
  \includegraphics[width=0.5 \textwidth]{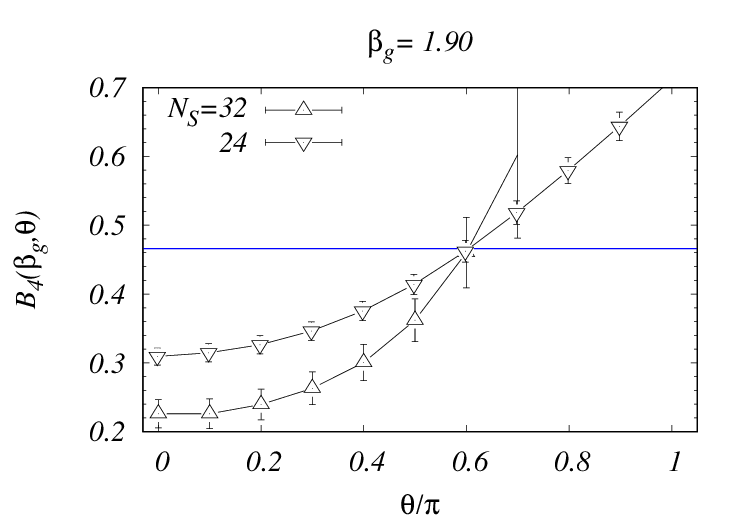}
  \end{tabular}
  \end{center}
 \caption{$\th$ dependence of $B_4$ at $\beta=1.91$ (left) and $1.90$ (right).
 The results from different lattices with $N_S=48$, 32, 24 are shown.
 }
 \label{fig:theta-bdc}
\end{figure}

The $\th$ dependence of $B_4(\beta_g,\th)$ at $\nape=0$ for $\beta_g=1.91$ and $1.90$ are shown in Fig.~\ref{fig:theta-bdc}, where the critical value $0.446$ is also shown by the horizontal line.
%
%
At $\beta=1.91$ (left panel), it is seen that the curves from different volumes intersect with each other around $B_4^{\rm crit}=0.466$ as they should do although the one with $N_S=48$ is noisy.
At $\beta=1.90$ (right panel), the result with $N_S=48$ is unavailable but the other two curves again intersect around the right point.
This coincidence of the critical $\th$ among different $N_S$ supports the validity of the analysis involving the extrapolation of sub-volume.
\begin{figure}[htb]
  \begin{center}
  \begin{tabular}{c}
  \includegraphics[width=0.7 \textwidth]{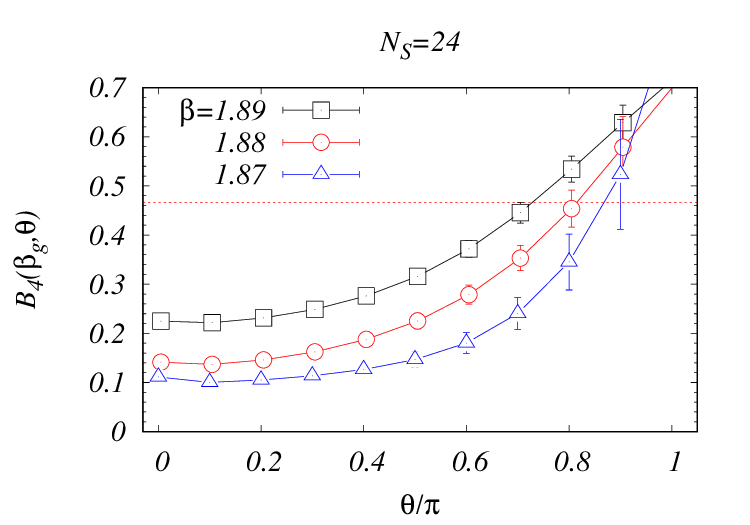}
  \end{tabular}
  \end{center}
 \caption{$\th$ dependence of $B_4$ obtained from the $N_S=24$ lattice.
 }
 \label{fig:theta-bdc-2}
\end{figure}

Below $\beta=1.90$, the results are available only from the $N_S=24$ lattices, which are plotted together in Fig.~\ref{fig:theta-bdc-2}.
It should be noted that the sub-volume method enables us to access $\th\sim 0.9\pi$ although further crosscheck is necessary.

\begin{figure}[t]
  \begin{center}
  \begin{tabular}{cc}
  \includegraphics[width=0.5 \textwidth]{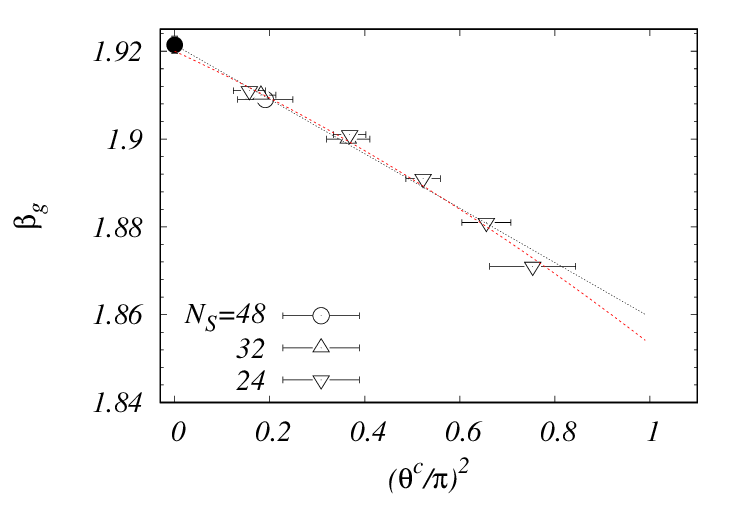}&
  \includegraphics[width=0.5 \textwidth]{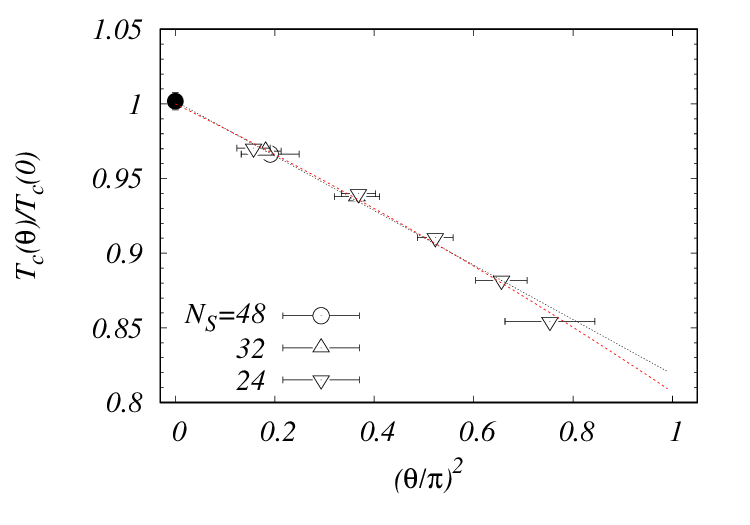}
  \end{tabular}
  \end{center}
 \caption{The critical curve in the $\th$-$\beta_g$ and $\th$-$T_c(\th)/T_c(0)$ planes.
 }
 \label{fig:theta-tc}
\end{figure}

We determine the critical value $\th^c$ of $\th$  at each $\beta_g$ by linearly interpolating two data points sandwiching $B_4^{\rm crit}=0.466$.
The resulting critical points are plotted in Fig.~\ref{fig:theta-tc} (left), which clearly shows that the critical $\beta_g$ decreases with $\th$.
Translating $\beta_g$ to $T/T_c(0)$ using the formula at $\th=0$ in~\cite{Giudice:2017dor}
(see also Fig.~\ref{fig:beta-vs-tovtc}), we obtain the phase boundary on the $\th$-$T/T_c(0)$ plane as shown in the right panel of Fig.~\ref{fig:theta-tc}.
We fit the data points in both plots to the functions,
\begin{align}
&   \beta_g^{\rm crit}(\th)
= c_0 + c_2\,\left({\th\over \pi}\right)^2
\label{eq:c0-c2}\,,\\
&   \beta_g^{\rm crit}(\th)
= 1 + d_2\,\left({\th\over \pi}\right)^2 + d_4\,\left({\th\over \pi}\right)^4
\label{eq:d2-d4}\,,\\
& {T_c(\th)\over T_c(0)}
= e_0 + e_2\,\left({\th\over \pi}\right)^2
\label{eq:e0-e2}\,,\\
& {T_c(\th)\over T_c(0)}
= 1 + f_2\,\left({\th\over \pi}\right)^2 + f_4\,\left({\th\over \pi}\right)^4
\label{eq:f2-f4}\,,
\end{align}
and the results are shown in the plots.

In order to guess the size of the systematic uncertainty associated with the quadratic extrapolation of the sub-volume, the linear extrapolations in $1/l$ are also performed, whose results at the intermediate step are shown in Figs.~\ref{fig:l-bdc-24}-\ref{fig:nape-bdc-48}.
Repeating the same analysis for those results, we determined the coefficients in \eqref{eq:c0-c2}-\eqref{eq:f2-f4} in this case.
By taking the difference as the systematic uncertainty, we finally obtain for the coefficients
\begin{align}
& c_0= 1.921(2)(1)\,,\quad
  c_2=-0.062(5)(4)\,,\quad
  (\chi^2/{\rm dof}=0.37)\\
& d_2= -0.026(4)(2)\,,\quad
  d_4=-0.009(7)(2)\,,\quad
  (\chi^2/{\rm dof}=0.001)\\
& e_0= 1.002(6)(0)\,,\quad
  e_2=-0.18(2)(1)\,,\quad
  (\chi^2/{\rm dof}=0.28)\\
& f_2=-0.16(2)(1)\,,\quad
  f_4=-0.03(4)(1)\,.\quad
  (\chi^2/{\rm dof}=0.01)\,,
\end{align}
where the first and second errors are statistical and systematic one, respectively, and the values of $\chi^2/$dof are for the quadratic fit.
Note that $c_0$ consistent with 1.92 are obtained without the input of $\beta_g(0)=1.92$.
This provides another nontrivial check for the validity of our analysis.

It is interesting to compare the $\th$ dependence with other \SU($N_c$) theories.
In~\cite{Bonanno:2023hhp}, the dependence is expressed as $T_c(\th)/T_c(0)=1-R\,\th^2+O(\th^4)$
for $N_c\ge 3$ with $R\sim 0.17/N_c^2$.
If one naively extrapolates this formula to $N_c=2$, we obtain $R\sim 0.04$.
Our result $R=e_2/\pi^2 \sim 0.02$ is clearly smaller than the naive extrapolation in $N_c$.
This is arguably not a surprise, since even the order of the phase transitions is different for the \SU$(2)$ theory as emphasized in the introduction.

\section{Further consistency check}
\label{sec:check}

As stated in the introduction, in the sub-volume method it is important to establish the validity of the extrapolation of the sub-volume.
In the previous section, we have seen that the numerical results indeed provide supports for the validity.
Here another consistency check is presented.

\begin{figure}[t]
  \begin{center}
  \begin{tabular}{cc}
  \includegraphics[width=0.5 \textwidth]{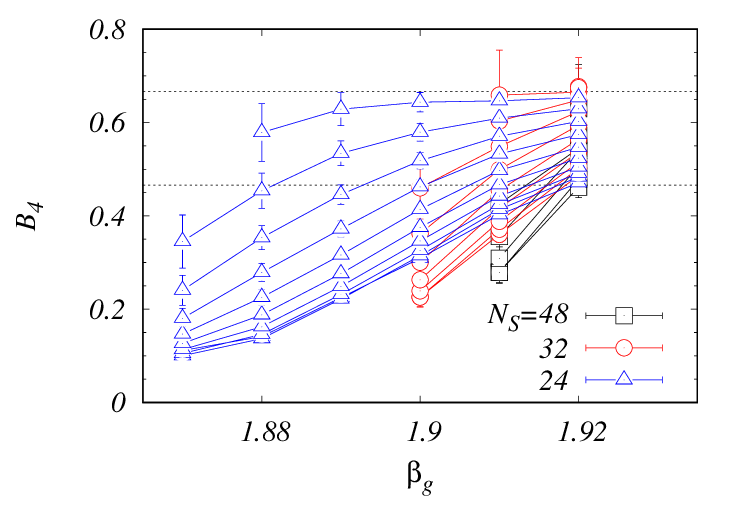}&
  \includegraphics[width=0.5 \textwidth]{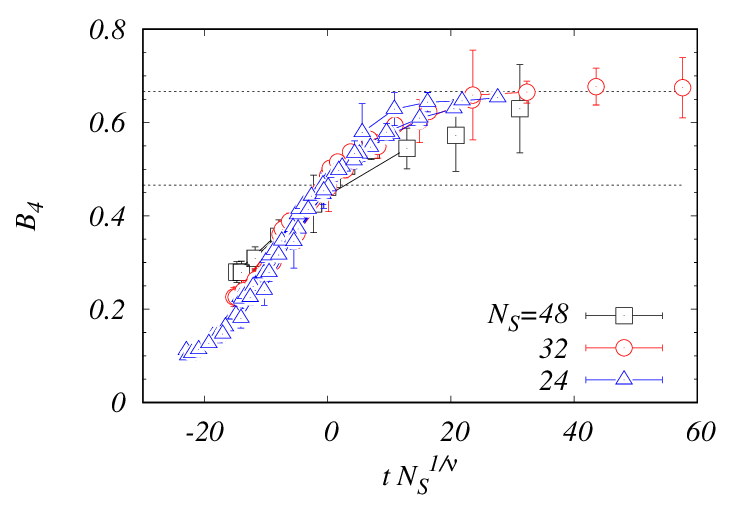}
  \end{tabular}
  \end{center}
 \caption{Test of the universality for $B_4$ for non-zero values of $\theta$.
The data having the same $\theta$ and $N_S$ are connected by a line.
 }
 \label{fig:univ-test}
\end{figure}

In Fig.~\ref{fig:scaling-plot}, we have seen that the rescaled Binder cumulants at $\theta=0$ collapse to a single scaling function independently of $N_S$.
We extend this to the case with $\theta\ne 0$.
In the left panel of Fig.~\ref{fig:univ-test}, the $\beta_g$ dependence of $B_4$ obtained by the extrapolation quadratic in $1/l$ are plotted for all available $\th$ and $N_S$.
The same data are re-plotted as a function of $tN_S^{1/\nu}$ on the right, where $\nu=0.6301$ for the 3d Ising model is used.
To calculate $t=T(\beta_g)/T(\beta_g^{\rm crit}(\theta))-1$ at arbitrary $\theta$ and $\beta_g$, we have used the fit result for \eqref{eq:d2-d4}.
It is seen that the widely scattered points on the left panel approximately form a single curve on the right.
We believe that this observation provides further support for the reliability of the extrapolation in $1/l$.

\section{Discussion}
\label{sec:discussion}

Following the argument by Keesom~\cite{Keesom:1932} and Ehrenfest~\cite{Ehrenfest:1933}\footnote{See also \cite{Buckingham:1961}.}
(for commentary, see, for example,~\cite{Jaeger:1998,Sauer:2016}), a curve of the second order phase
transition in phase diagram is expressed in terms of partial derivatives of the free energy density.
This argument, which is to be described below, is a generalization of the Clausius-Clapeyron relation for the first order phase transition.

We define the free energy densities $F_{H,L}(T,\th)=-\ln Z(T,\th)/V_3$, where $F_H$ and $F_L$ are defined
only in the high ($T\ge T_c(\th)$) and  low ($T\le T_c(\th)$) temperature phase, respectively, and analytic
in the phases they are defined except on the phase boundary.
On the $\th$-$T$ plane, consider two points, $A=(0,T_c(0))$ and $B=(\delta\th,T_c(\delta\th))$,
which are on the critical curve.
When the two points are very close to each other, the critical curve between them
can be approximated by a line.
Using the CP symmetry $T_c(\th)=T_c(-\th)$ and assuming that $T_c(\th)$ is regular around $\th=0$,
we express $\delta T_c=T_c(\delta\th)-T_c(0)$ as
\begin{align}
   {\delta T_c\over T_c(0)}
=  {\tilde f}_2 (\delta\th)^2 + {\tilde f}_4  (\delta\th)^4 + \cdots
 \,,
 \label{eq:deltaTc}
\end{align}
where ${\tilde f}_2=f_2/\pi^2$ and ${\tilde f}_4=f_4/\pi^4$ are related to the coefficients in \eqref{eq:f2-f4}.
Since the phase transition under consideration is of second order, the following equalities hold
at $A$ and $B$ on the curve,
\begin{align}
& \lim_{\epsilon\to 0}{F_H\big|_{A+\epsilon}\over F_L\big|_{A-\epsilon}}
= \lim_{\epsilon\to 0}{F_H\big|_{B+\epsilon}\over F_L\big|_{B-\epsilon}}
= 1\,,
 \label{eq:condition-1}\\
& \lim_{\epsilon\to 0}
  {{\pde F_H \over \pde T}\big|_{A+\epsilon}\over
   {\pde F_L \over \pde T}\big|_{A-\epsilon}}
= \lim_{\epsilon\to 0}
  {{\pde F_H \over \pde T}\big|_{B+\epsilon}\over
   {\pde F_L \over \pde T}\big|_{B-\epsilon}}
=1\,.
 \label{eq:condition-3}
\end{align}
where
\begin{align}
 A\pm\epsilon=(0,\,T_c(0)\pm\epsilon),
\qquad
 B\pm\epsilon=(\delta\th,\,T_c(\delta\th)\pm\epsilon)\,,
\end{align}
and the derivatives are taken where the functions are analytic.
Introducing the notation,
\begin{align}
&  F_{H,L}^{(l,m)}(A\pm\epsilon)
= {\partial^{l+m}F_{H,L}(\th,T)\over \partial T^l\,\partial \th^m}
 \bigg|_{A\pm\epsilon}
 \,,
\end{align}
we rewrite \eqref{eq:condition-3} as
\begin{align}
& \lim_{\epsilon\to 0}\bigg[
  \left(F_H^{(1,0)}(B+\epsilon)-F_H^{(1,0)}(A+\epsilon)\right)
- \left(F_L^{(1,0)}(B-\epsilon)-F_L^{(1,0)}(A-\epsilon)\right)
 \bigg]= 0
\label{eq:eq-2ndpt}
\,.
\end{align}
Performing the Taylor expansion in each parenthesis around $A\pm\epsilon$ along the critical line in \eqref{eq:deltaTc} and collecting $O((\delta\th)^2)$ terms, one finds that ${\tilde f}_2$ is expressed as
\begin{align}
   {\tilde f}_2
&=-{1\over 2 T_c(0)}
   \lim_{\epsilon\to 0+}{\Delta F_{HL}^{(1,2)}\over\Delta F_{HL}^{(2,0)}}
\,,
\label{eq:c2_formula}
\end{align} 
where we have defined
\begin{align}
& \Delta F_{HL}^{(l,m)}
= F_H^{(l,m)}(A+\epsilon) - F_L^{(l,m)}(A-\epsilon)
\,.
\end{align}
$F_{H,L}^{(2,0)}$ are the specific heat, which is known to diverge as
$|\epsilon|^{-\alpha}A_{H,L}$, respectively, from the scaling analysis.
The exponent and the universal ratio of the amplitudes are estimated to be
$\alpha=0.110(1)$~\cite{Pelissetto:2000ek} and $A_L/A_H\sim 0.55$~\cite{Hasenbusch:1997be}
in the 3d Ising model, respectively.
On the other hand, $F_{H,L}^{(1,2)}\sim \partial \chi_t/\partial T$, where $\chi_t=\langle Q^2 \rangle\,T/V_3$ is the topological susceptibility.
Since we have observed in the numerical simulation that ${\tilde f}_2=f_2/\pi^2$ is finite,
$\partial \chi_t/\partial T$ has to diverge with the same exponent as the specific heat, {\it i.e.}
\begin{align}
  {\partial \chi_t\over \partial T}\bigg|_{A+\epsilon}
- {\partial \chi_t\over \partial T}\bigg|_{A-\epsilon}
\propto |\epsilon|^{-\alpha}
\label{eq:chit-singularity}
\,.
\end{align}
Indeed, for an ansatz to satisfy the universality, i.e., $F_H - F_L \propto |t|^{2-\alpha}$ with $t = T/T_c(\theta) - 1$, one can explicitly see that $\Delta F_{HL}^{(1,2)}$ and $\Delta F_{HL}^{(2,0)}$ in \eqref{eq:c2_formula} both scale as $|\epsilon|^{-\alpha}$.
In~\cite{Yamada:2024vsk}, we have calculated $\chi_t$ around $T_c$, and determined its overall $T$ dependence.
However, it is yet difficult to numerically establish the presence of the singularity predicted in~\eqref{eq:chit-singularity}, because of the lack of accuracy and the smallness of
the exponent $\alpha\sim 0.1$.
A drastic improvement in accuracy seems necessary to confirm the singularity,
and we leave this task for future works.

\section{Summary}
\label{sec:summary}

We have explored the $\th$-$T$ phase diagram in 4d \SU(2) pure YM theory using lattice numerical simulations.
In order to mitigate the sign problem due to finite $\th$, we proposed the method in which the re-weighting is applied only to a sub-volume at the price of introducing the extrapolation of the sub-volume.
With this method, we determined the fourth-order Binder cumulant of the order parameter at finite values of $\th$ to locate the critical point.
With the full use of the universality, the curve of confinement-deconfinement transition on the phase diagram was determined up to $\th\sim 0.9\pi$ at $N_T=8$.
Defining $T_c(\th)/T_c(0)=1-R\,\th^2+O(\th^4)$, our result $R\sim 0.017$ is clearly smaller than the extrapolation of the large $N_c$ fit to $N_c=2$ ($R\sim 0.17/N_c^2$), and this seems to be consistent with the peculiarity of the \SU$(2)$ theory that the transition is of second order.
We have estimated the possible size of the systematic uncertainty associated with the sub-volume extrapolation by performing alternative method.
In the sub-volume method, it is important to establish the reliability of the extrapolation.
We have confirmed that the results obtained with the extrapolation show consistency with the properties of the Binder cumulant at the critical point and the second order phase transition.
Following the argument by Keesom and Ehrenfest, we have shown that $\chi_t$ has to have a singular $T$ dependence at $T_c$, which we hope to confirm numerically in the future.

In this paper, we focused on the feasibility study of the methods developed here. 
Further studies on systematic errors and the continuum extrapolation would be the next step.
Our method of the re-weighting on sub-volume does not rely on expansion nor analytic continuation in $\th$.
It is interesting to see how the method works in the search for the critical endpoint in the $\mu$-$T$ phase diagram in the finite-density QCD.

\section*{Acknowledgments}
This work is based in part on the Bridge++ code~\cite{Ueda:2014rya}
and is supported in part by JSPS KAKENHI Grant-in-Aid for Scientific
Research (Nos.~19H00689 [RK, NY, MY], 21H01086, 23K20847, 22K21350~[RK], 
22K03645 [NY], 19K03820, 20H05860, 23H01168 [MY]), and JST, Japan
(PRESTO Grant No.~JPMJPR225A, Moonshot R\&~D Grant No.~JPMJMS2061 [MY]).
This research used computational resources of Cygnus and
Wisteria/BDEC-01 Odyssey (the University of Tokyo), provided by the
Multidisciplinary Cooperative Research Program in the Center for
Computational Sciences, University of Tsukuba.


\end{document}